\documentclass[twocolumn]{aastex7}

\usepackage{amsmath}

\shorttitle{Testing SIDM with SDSS J0946+1006}
\shortauthors{Li et al.}
\graphicspath{{./}}

\begin{document}

\title{The ``Little Dark Dot'': Evidence for Self-interacting Dark Matter\\ in the Strong Lens SDSS J0946+1006?}

\author[0009-0004-0904-7400]{Shubo Li}
\affiliation{School of Physics and Astronomy, Beijing Normal University, Beijing 100875, China}
\affiliation{National Astronomical Observatories, Chinese Academy of Sciences, 20A Datun Road, Chaoyang District, Beijing 100101, China}
\affiliation{School of Astronomy and Space Science, University of Chinese Academy of Sciences, Beijing 100049, China}
\email{lisb@nao.cas.cn}

\author[0000-0003-3899-0612]{Ran Li}
\affiliation{School of Physics and Astronomy, Beijing Normal University, Beijing 100875, China}
\affiliation{School of Astronomy and Space Science, University of Chinese Academy of Sciences, Beijing 100049, China} 
\email{liran@bnu.edu.cn}

\author[0009-0005-3671-1861]{Kaihao Wang}
\affiliation{School of Physics and Astronomy, Beijing Normal University, Beijing 100875, China}
\affiliation{National Astronomical Observatories, Chinese Academy of Sciences, 20A Datun Road, Chaoyang District, Beijing 100101, China}
\affiliation{School of Astronomy and Space Science, University of Chinese Academy of Sciences, Beijing 100049, China}
\email{wangkh@nao.cas.cn}

\author[0009-0008-2362-1367]{Zixiang Jia}
\affiliation{Department of Astronomy, Peking University, Beijing 100871, China}
\email{jiazixiang@stu.pku.edu.cn}

\author[0000-0003-4988-9296]{Xiaoyue Cao}
\affiliation{School of Astronomy and Space Science, University of Chinese Academy of Sciences, Beijing 100049, China}
\affiliation{National Astronomical Observatories, Chinese Academy of Sciences, 20A Datun Road, Chaoyang District, Beijing 100101, China}
\affiliation{School of Physics and Astronomy, Beijing Normal University, Beijing 100875, China}
\email{xycao@nao.cas.cn}

\author[0000-0002-2338-716X]{Carlos S. Frenk}
\affiliation{Institute for Computational Cosmology, Department of Physics, Durham University, South Road, Durham DH1 3LE, UK}
\email{c.s.frenk@durham.ac.uk}

\author[0000-0001-6115-0633]{Fangzhou Jiang}
\affiliation{Kavli Institute for Astronomy and Astrophysics, Peking University, Beijing 100871, China}
\email{fangzhou.jiang@pku.edu.cn}

\author[0000-0002-4465-1564]{Aristeidis Amvrosiadis}
\affiliation{Institute for Computational Cosmology, Department of Physics, Durham University, South Road, Durham DH1 3LE, UK}
\email{aristeidis.amvrosiadis@durham.ac.uk}

\author[0000-0002-5954-7903]{Shaun Cole}
\affiliation{Institute for Computational Cosmology, Department of Physics, Durham University, South Road, Durham DH1 3LE, UK}
\email{shaun.cole@durham.ac.uk}

\author[0000-0003-3672-9365]{Qiuhan He}
\affiliation{Institute for Computational Cosmology, Department of Physics, Durham University, South Road, Durham DH1 3LE, UK}
\email{qiuhan.he@durham.ac.uk}

\author[0009-0007-0679-818X]{Samuel C. Lange}
\affiliation{Institute for Computational Cosmology, Department of Physics, Durham University, South Road, Durham DH1 3LE, UK}
\email{samuel.c.lange@durham.ac.uk}

\author[0000-0002-6085-3780]{Richard Massey}
\affiliation{Institute for Computational Cosmology, Department of Physics, Durham University, South Road, Durham DH1 3LE, UK}
\email{r.j.massey@durham.ac.uk}

\author[0000-0002-8987-7401]{James W. Nightingale}
\affiliation{School of Mathematics, Statistics and Physics, Newcastle University, Newcastle upon Tyne, NE1 7RU, UK}
\affiliation{Institute for Computational Cosmology, Department of Physics, Durham University, South Road, Durham DH1 3LE, UK}
\affiliation{Centre for Extragalactic Astronomy, Department of Physics, Durham University, South Road, Durham DH1 3LE, UK}
\email{James.Nightingale@newcastle.ac.uk}

\author[0000-0002-0086-0524]{Andrew Robertson}
\affiliation{Carnegie Observatories, 813 Santa Barbara Street, Pasadena, CA 91101, USA}
\email{arobertson@carnegiescience.edu}

\author[0000-0003-4986-5091]{Maximilian von Wietersheim-Kramsta}
\affiliation{Institute for Computational Cosmology, Department of Physics, Durham University, South Road, Durham DH1 3LE, UK}
\affiliation{Centre for Extragalactic Astronomy, Department of Physics, Durham University, South Road, Durham DH1 3LE, UK}
\email{maximilian.von-wietersheim-kramsta@durham.ac.uk}

\author{Xianghao Ma}
\affiliation{School of Physics and Astronomy, Beijing Normal University, Beijing 100875, China}
\affiliation{National Astronomical Observatories, Chinese Academy of Sciences, 20A Datun Road, Chaoyang District, Beijing 100101, China}
\affiliation{School of Astronomy and Space Science, University of Chinese Academy of Sciences, Beijing 100049, China}
\email{maxh@nao.cas.cn}

\correspondingauthor{Ran Li}
\email{liran@bnu.edu.cn}

\begin{abstract}
Previous studies, based on precise modeling of a gravitationally lensing image, have identified what may be an extremely compact, dark perturber in the well-known lensing system SDSS J0946+1006 (the ``Jackpot''). Its remarkable compactness challenges the standard cold dark matter (CDM) paradigm. In this paper, we explore whether such a compact perturber could be explained as a core-collapse halo described by the self-interacting dark matter (SIDM) model. Using the isothermal Jeans method, we compute the density profiles of core-collapse halos across a range of masses. Our comparison with observations indicates that a core-collapse halo has an inner density profile and mass enclosed within 1 kpc that fit the data well, but only if the halo has a total mass $\sim10^{11}~{\rm M_{\odot}}$. While a halo of this mass should host a detectable galaxy, the current observational upper limit on the perturber's luminosity remains uncertain. Resolving whether or not the data support the presence of a core-collapse SIDM halo therefore requires future deep observations to measure its luminosity.
\end{abstract}

\section{Introduction} \label{sec:introduction}
The standard theory of cosmic structure formation is built upon the cold dark matter (CDM) model. In this framework, structures grow hierarchically: smaller dark matter halos form first and later merge and accrete material to evolve into larger systems. Galaxies form within these halos and coevolve with their hosts. A direct prediction of this paradigm is the existence of a large number of low-mass dark matter halos in the Universe today.  

Detecting these predicted CDM halos is challenging. These halos can host dwarf galaxies, which serve as tracers, but in the distant Universe, even if dwarf galaxies are observed, measuring the masses of their host halos is often difficult. Moreover, baryonic feedback processes are expected to suppress star formation in halos smaller than $10^9~{\rm M_\odot}$ \citep[e.g.,][]{Sawala2015, Sawala2016, Bullock2017}, leaving many low-mass halos either completely dark or hosting faint galaxies that are hard to detect. 

Beyond the Milky Way, strong gravitational lensing provides the only method currently available to directly detect these dark halos and measure their masses and internal structures. Such halos, if present, can perturb lensing systems by introducing anomalies in flux ratios or causing subtle distortions in lensing images \citep{Mao1998, Vegetti2009}. These dark halos may reside within the main lens galaxy or along the line of sight to the background source \citep{Li_cdmwdm2016, Li2017_los, Despali2018, Gilman2018, He2022_los, He2022_abc}.  

Over the past decade, several candidate dark halos have been identified in high-resolution imaging \citep{Vegetti_2010, Vegetti2012, Hezaveh2016_alma_subhalo}. Among these, SDSS J0946+1006 stands out as particularly intriguing. This strong-lensing system was first discovered in the Sloan Digital Sky Survey main spectroscopic sample by the SLACS survey and subsequently received deep follow-up observations with the Hubble Space Telescope (HST), accumulating a total exposure time of 2096 s \citep[$I$-band;][]{Gavazzi_2008}. It has become one of the most extensively studied systems, with multiple analyses consistently identifying evidence for a dark perturber \citep{Vegetti_2010, Minor2021, Ballard_24, Enzi2024, Nightingale_24, Despali2024}. Using pixelized image modeling of the lensing data, \citet{Vegetti_2010} reported the detection of a dark perturber in this system. Fitting the perturber with a tidally truncated pseudo-Jaffe profile, characterized by an inner density profile scaling as \(\rho \sim r^{-2}\), they estimated a subhalo mass of $3.51\times10^9~{\rm M_\odot}$.  

However, in a CDM Universe, pure dark matter halos are expected to follow a Navarro--Frenk--White (NFW) profile \citep{NFW1997}, characterized by an inner density profile scaling as \(\rho \sim r^{-1}\), which is much shallower than that of a pseudo-Jaffe profile. \citet{Despali2018} demonstrated that fitting an NFW perturber with a pseudo-Jaffe model could lead to halo mass underestimation by an order of magnitude; they corrected the perturber’s mass to $2.06\times10^{10}~{\rm M_\odot}$. Subsequently, \citet{Minor2021} adopted a truncated NFW (tNFW) model, allowing the concentration parameter to vary freely. They found that the perturber is exceptionally compact, with a subhalo mass of $5.01\times10^9~{\rm M_\odot}$ and an extremely high concentration of 1560. The inferred parameters become less extreme when ``multipoles'' are included in the macro model, suggesting deviations from ellipticity in the primary lens. The best-fit values are $M_{200} = 3.02\times10^{10}~{\rm M_\odot}$ and $c_{200}=70.6$, which still remain at least $5 \sigma$ above the median concentration of CDM field halos derived from $N$-body cosmological simulations. These findings have been corroborated by subsequent studies \citep{Enzi2024, Despali2024}.
  
It is worth noting that the perturber's compactness cannot be easily explained by assuming it lies along the line of sight rather than in the lens plane. \citet{He2022_los} argued that the perturbation induced by a dark halo is strongest when it resides in the lens plane. For a perturber along the line of sight to produce comparable distortions, it would require a higher mass or a steeper density profile. \citet{Enzi2024} allowed the perturber’s redshift to vary in their analysis of the system and concluded that the subhalo is more likely to reside in the lens plane.  

The unusual steepness of this halo has prompted consideration of alternative dark matter models, particularly self-interacting dark matter (SIDM), which has been proposed as a potential revision to CDM on small scales \citep{Spergel2000, Tulin&Yu}. In a self-gravitating system, the behavior of dark matter exhibits similarities to that of stars in a globular cluster, and the evolution of an SIDM halo ultimately leads to the ``gravothermal catastrophe'' \citep{Lynden&Bell1968}: over sufficiently long timescales, the self-interaction-driven dark matter core becomes thermally unstable. This instability leads to gravitational collapse, ultimately reestablishing a cuspy density profile in the halo center \citep[e.g.,][]{Balberg2002PRL, Balberg2002ApJ, Pollack2015, Essig2019, Nishikawa2020}. Drawing on the first group-scale $N$-body simulation with large-amplitude, velocity-dependent dark matter self-interactions, \citet{Nadler2023} demonstrated that SIDM can produce halos with properties similar to the J0946+1006 perturber. However, comprehensive studies on whether SIDM core-collapse halos can explain the dark perturber in J0946 remain limited, as most previous analyses have been qualitative rather than quantitative \citep[e.g.,][]{Nadler2023, Enzi2024}.

In this paper, we systematically investigate the possibility that the ``little dark dot'' in J0946+1006 is a core-collapse SIDM halo. We compute the density profiles using an improved version of the isothermal Jeans model that can capture the onset of core collapse, across a wide range of halo mass and compare the model predictions with the lensing data. Additionally, we generate mock lensing images with the same configuration as J0946+1006, incorporating a theoretical SIDM halo profile to simulate image distortions. By fitting the distortions, we evaluate the biases introduced during modeling. Our results indicate that an SIDM halo with a mass of $10^{11}~{\rm M_\odot}$ is required to produce the observed lensing perturbation. However, this raises a new question: why does such a massive halo not host a prominent visible galaxy?

This paper is organized as follows. Section~\ref{sec:Methods} presents our methodology for modeling SIDM halos ($\S$~\ref{sec:method_density_profile}), simulating their evolutionary processes ($\S$~\ref{sec:method_full_evolution}), and computing their lensing properties ($\S$~\ref{sec:2d_properties}). In Section~\ref{sec:Results}, we compare the results of our SIDM modeling with observational results ($\S$~\ref{sec:Results_1}) and evaluate the performance of gravitational lens fitting using the mock lensing image ($\S$~\ref{sec:Results_2}). Section~\ref{sec:Discussion} presents a discussion of our results, focusing on their implications and limitations, and finally, Section~\ref{sec:Conclusions} summarizes the principal conclusions of this study and outlines potential directions for future research.

Throughout this paper, we adopt the Planck ${\rm \Lambda CDM}$ cosmology \citep{Planck2015} with $\Omega_m = 0.307$ and $H_0 = 67.7~{\rm km}~{\rm s}^{-1}~{\rm Mpc}^{-1}$.

\section{Methods} \label{sec:Methods}
\subsection{Solving the density profile of the SIDM halo}
\label{sec:method_density_profile}
Considering its low computational cost and the convenience of being able to model specific galaxies and dark matter halo systems, we use a semi-analytical method called isothermal Jeans method to construct the SIDM halo. This method was first proposed by \cite{Kaplinghat2014, Kaplinghat2016} and its efficacy for SIDM density profiles has been intensively tested using hydrodynamical simulations of SIDM \citep[e.g.,][]{Robertson2021}. Here, we present the basic idea and implementation process of this method. For more detailed information, readers are referred to \citet{Jiang2023}, where a public code implementation is also provided.

The model is based on the idea that energy transfer through elastic scattering between dark matter particles leads to the formation of an isothermal SIDM core after several interactions over the halo's evolutionary timescale. The density of this isothermal core $\rho_{\rm iso}$ should be describable by the Jeans equation,
\begin{equation}
    \frac{\mathrm{d}(\rho_{\rm iso}\nu^2)}{\mathrm{d}r}+\frac{2\beta}{r}\nu^2\rho_{\rm iso}=-\rho_{\rm iso}\frac{\mathrm{d}\Phi}{\mathrm{d}r},
	\label{eq:JeansEquation}
\end{equation}
where $\Phi$ is the total gravitational potential, including both dark matter and baryons. Assuming an isotropic ($\beta=0$) and constant velocity dispersion ($\nu=\nu_0$) in the core region, we get a simple generic solution of the Jeans equation,
\begin{equation}
    \rho_{\rm iso}(r) = \rho_{\rm 0} \exp\left(-\frac{\Delta\Phi}{\nu_0^2}\right),
	\label{eq:rho_isothermal}
\end{equation}
where $\rho_{\rm 0}$ is the central density of the isothermal core and $\Delta\Phi=\Phi(r)-\Phi(0)$. To obtain the expression of $\Phi(r)$, we apply the Poisson equation, which connects the gravitational potential to the matter density,
\begin{equation}
    \frac{1}{r^2} \frac{\mathrm{d}}{\mathrm{d}r} \left( r^2 \frac{\mathrm{d}\Phi}{\mathrm{d}r} \right) = 4\pi G \left( \rho_{\rm iso} + \rho_{\rm b} \right),
	\label{eq:PoissonEquation}
\end{equation}
where $\rho_{\rm b}$ represents the density distribution of the galaxy. The dark matter density (Eq.~\ref{eq:rho_isothermal}) and the galaxy's density are then substituted into the Poisson equation (Eq.~\ref{eq:PoissonEquation}). By applying appropriate variable transformations and setting the boundary conditions, the resulting second-order differential equation is solved to obtain $\rho_{\rm iso}$ as a function of radius. More details for each step can be found in Appendix~\ref{sec:appendix_poisson_equation}. 

For a given set of $\rho_{\rm 0}$ and $\nu_0$, we can determine the density profile of the isothermal core of the SIDM halo following the above steps. However, elastic scattering between dark matter particles primarily occurs in the central regions, where the particle density is sufficiently high. In the outer regions of the dark halo, the interaction between particles becomes negligible, and the dark matter distribution still follows the predictions of CDM, i.e.,\ the NFW profile:
\begin{equation}
    \rho_{\rm NFW}(r) = \frac{\rho_s}{\frac{r}{r_s}\left(1+\frac{r}{r_s}\right)^2}.
    \label{NFW}
\end{equation}

After defining $r_{200}$ as the radius within which the mean enclosed density is 200 times the critical density $\rho_{\rm crit}$, and $M_{200}$ as the mass within $r_{200}$, the scaling parameters $\rho_s$ and $r_s$ in Eq.~\ref{NFW} can be replaced with the halo mass $M_{200}$ and the concentration parameter $c_{200}=r_{200}/r_s$. Now the SIDM halo is divided into two regions, and we need to identify the boundary that separates them. The stitching radius $r_1$ is defined as the radius at which the local scattering rate $\Gamma(r)$, multiplied by the halo's lifetime $t_{\rm age}$, equals 1, marking the transition point where the behavior of the dark halo changes:
\begin{equation}
    \begin{aligned}
        \Gamma(r_1) &= \langle v_{\rm pair}(r_1)\rangle \rho_{\rm dm}(r_1)\sigma_m \\
        &= \frac{4}{\sqrt{\pi}} \nu(r_1) \rho_{\rm dm}(r_1) \sigma_m = \frac{1}{t_{\rm age}}.
    \end{aligned}
    \label{eq:r1}
\end{equation}

\begin{figure*}
    \centering
	\includegraphics[width= \textwidth]{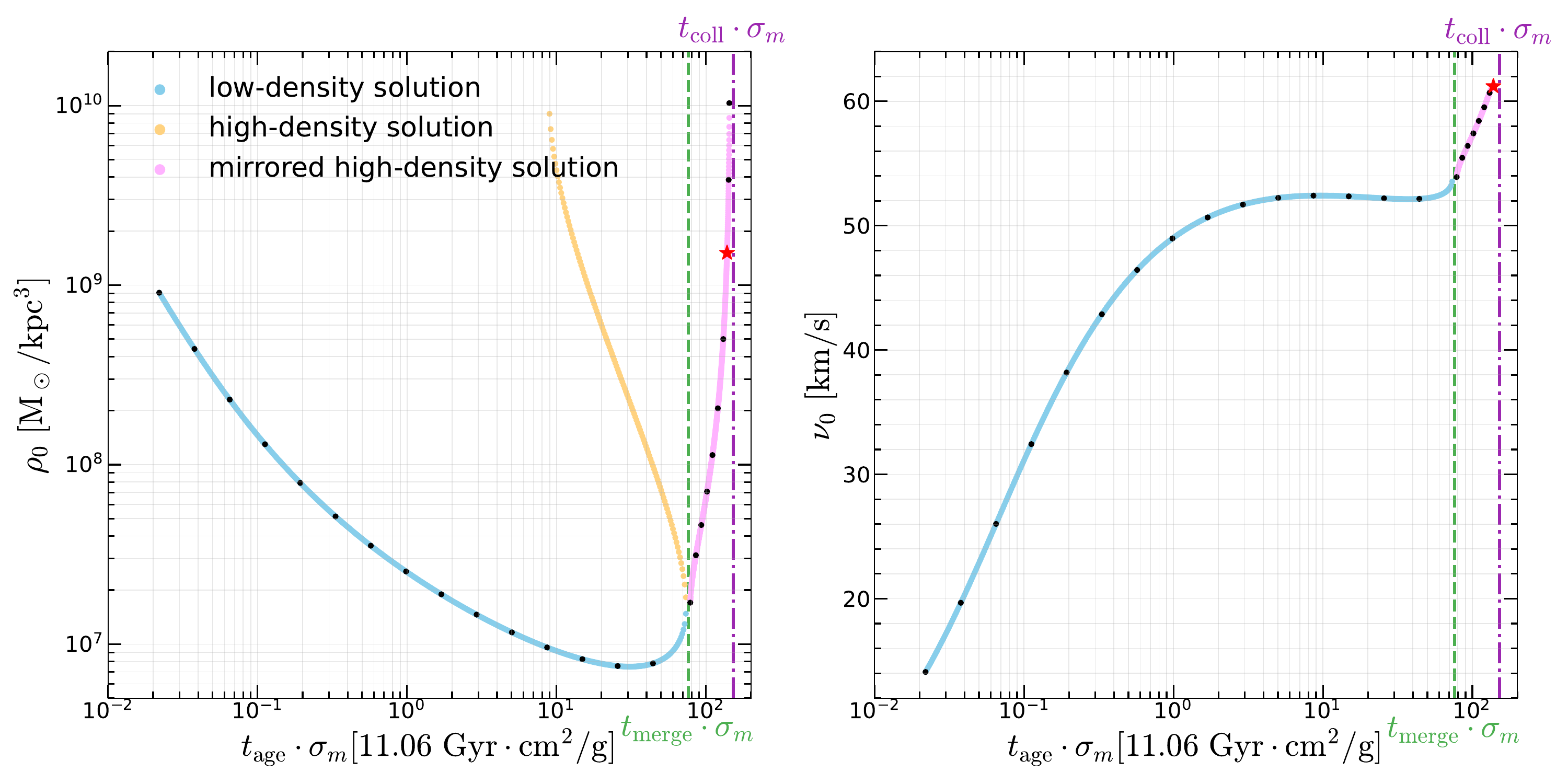}
    \caption{The temporal evolution of the central density $\rho_0$ (left panel) and velocity dispersion $\nu_0$ (right panel) of an SIDM core with $M_{200}=1\times10^{11}~{\rm M_\odot}$. Considering the degeneracy between the formation time $t_{\rm age}$ and the scattering cross section $\sigma_{m}$, we choose their product as the horizontal axis, with time measured in units of the cosmic time of J0946+1006 (11.06 Gyr). The blue, orange, and magenta scatter points represent the evolution for the low-density solution, high-density solution, and mirrored high-density solution, respectively. The moments corresponding to $t_{\rm merge}$ and $t_{\rm coll}$ are marked by a green dashed line and a purple dashed-dotted line, respectively. The black dots represent the moments we chose to examine the evolution of the SIDM profile, while the red star represents the moment closest to the result of \citet{Minor2021} in the 2D observational space.}
    \label{fig:evolution}
\end{figure*}

Here, $\sigma_m$ denotes the effective self-interaction cross section per particle mass (in units of ${\rm cm^2/g}$), which we treat as a constant following \citet{Yang_Yu2022}'s framework. This approximation replaces the velocity-dependent cross section with a single characteristic value for each halo. $\rho_{\rm dm}$ is the dark matter density profile, which can be set either by the inner isothermal core or by the CDM outskirts. $\langle v_{\rm pair}(r)\rangle$ denotes the mean pairwise velocity of particles at radius $r$, and for a Maxwell–Boltzmann velocity distribution with a 1D velocity dispersion profile $\nu(r)$, it is given by $\langle v_{\rm pair}(r)\rangle = \frac{4}{\sqrt{\pi}}\nu(r)$. 

In practice, a halo's $t_{\rm age}$ is difficult to precisely define in a cosmology where structures grow hierarchically. A common operational definition is the duration required for the main progenitor to grow from $M_{200}/2$ to $M_{200}$. Within our forward-modeling framework, the SIDM density profile becomes uniquely determined by the product $t_{\rm age} \cdot \sigma_m$. Therefore, in this work, when comparing with observations, we compute a series of density profiles for halos with a given mass under varying values of $t_{\rm age} \cdot \sigma_m$.

Once $r_1$ is determined, we can stitch these two regions together by imposing the matching conditions, $\rho_{\rm iso}(r_1) \approx \rho_{\rm cdm}(r_1)$ and $M_{\rm iso}(r_1) \approx M_{\rm cdm}(r_1)$, and then minimizing the ``stitching error'' term $\delta^2$:
\begin{equation}
    \delta^2 = \left[\frac{\rho_{\rm iso}(r_1)-\rho_{\rm cdm}(r_1)}{\rho_{\rm cdm}(r_1)}\right]^2+\left[\frac{M_{\rm iso}(r_1)-M_{\rm cdm}(r_1)}{M_{\rm cdm}(r_1)}\right]^2.
	\label{eq:delta}
\end{equation}

Here, we adopt the \textit{outside-in} approach \citep{Robertson2021, Sagunski2021}. For a CDM halo with known $M_{200}$, $c_{200}$, and the inner galaxy's density distribution, we first compute the NFW profile for its outskirts. For a given $t_{\rm age}\cdot\sigma_m$ value, we determine $r_1$ using Eq.~\ref{eq:r1}. With a set of ($\rho_0$, $\nu_0$) values, we then derive the isothermal core density profile by solving the coupled Jeans and Poisson equations (Eq.~\ref{eq:JeansEquation} and Eq.~\ref{eq:PoissonEquation}). Finally, we optimize the ($\rho_0$, $\nu_0$) combination to minimize $\delta^2$ (Eq.~\ref{eq:delta}), ensuring a smooth transition between the isothermal core and CDM outskirts at $r_1$. In summary, once $M_{200}$, $c_{200}$, $t_{\rm age}$, $\sigma_m$, and the galaxy’s density profile are provided, this isothermal Jeans model yields the complete density profile of the SIDM halo.

In Appendix~\ref{sec:appendix_poisson_equation}, we discuss the impact of baryonic matter on SIDM halos within our SIDM model. Given the nondetection of an optical counterpart for J0946+1006 and the upper limit of its luminosity should be no bigger than $5\times10^6~{\rm L_\odot}$ \citep{Vegetti_2010}, we restrict our analysis exclusively to the dark-matter-only situation. Specifically, we exclude the galactic density profile from Eq.~\ref{eq:PoissonEquation} and employ the NFW profile alone to derive the density profile of the inner SIDM core.

\subsection{Computing the full time evolution of the SIDM halo}
\label{sec:method_full_evolution}

\begin{figure*}
	\includegraphics[width=\textwidth]{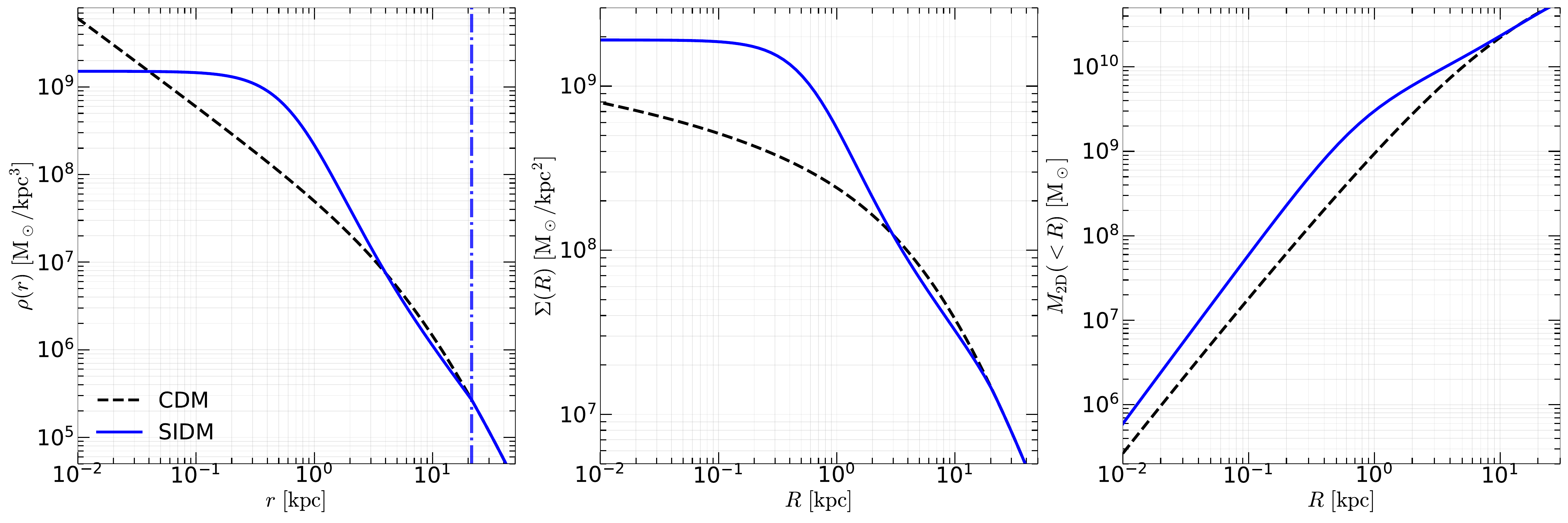}
    \caption{The density and mass profiles of a core-collapse SIDM halo with $M_{200}=10^{11}~{\rm M_\odot}$ (corresponding to the evolutionary stage indicated by the red star in Fig.~\ref{fig:evolution}) and its CDM counterpart. SIDM and CDM profiles are shown in blue and black, respectively. Left: 3D density profiles, with the blue vertical dashed-dotted line marking the stitching radius, $r_1$. Middle: 2D projected surface density profiles. Right: 2D enclosed mass profiles.}
    \label{fig:profiles1}
\end{figure*}

In an SIDM halo, heat transport driven by collisions between dark matter particles governs the temporal evolution of halo internal structures. Initially, these collisions facilitate heat flow inward along a positive temperature gradient, leading to the formation of a central core. This process, known as \textit{core formation}, results in the gradual growth of the core as the inner region undergoes thermalization. Once thermal contact extends to the outskirts of the halo, heat begins to flow outward, marking the onset of the \textit{core-collapse} phase. Due to the negative specific heat characteristic, the core area becomes hotter and denser after losing heat, leading the temperature and density of the isothermal core at the halo center increase steadily, culminating in a gravothermal catastrophe \citep{Tulin&Yu}.

The full time evolution of an SIDM halo can be more rigorously captured using the gravothermal fluid formalism, which models SIDM halos by solving a set of coupled partial differential equations for matter and heat transport with a CDM halo as the initial condition \citep[e.g.,][]{Essig2019, Nishikawa2020, Yang2023}. In contrast, the isothermal Jeans method assumes that the inner region is in approximate equilibrium, an assumption that breaks down during the core-collapse phase, making the standard approach inadequate for capturing the behavior of SIDM halos in this regime.
 
However, \citet{Jiang2023} noted that prior to core collapse, the isothermal Jeans method allows for both high-density and low-density solutions for the isothermal core: during the minimization of $\delta^2$, two sets of solutions $(\rho_0, \nu_0)$ are found, where the $\nu_0$ values are close to each other, but the $\rho_0$ values differ by several orders of magnitude (see Figure~1 of \citealt{Jiang2023}). These two sets of solutions gradually converge at a specific time, denoted as $t_{\rm merge}$. 

Based on this, \citeauthor{Yang2024}~(\citeyear{Yang2024}, inspired by private communications with M. Kaplinghat) demonstrated that mirroring the high-density solution at the merger time $t_{\rm merge}$ yields isothermal solutions corresponding to the core-collapse phase. This leads to a natural definition of the collapse timescale as $t_{\rm coll}=2\times t_{\rm merge}$. As shown in Figure 2 of \citet{Yang2024}, although minor discrepancies exist, the gravothermal and isothermal solutions exhibit remarkable similarity throughout the entire halo evolution. The physical justification for this mirroring approach can be understood as follows. Since time is treated as a scalar quantity in Eq.~\ref{eq:r1}, the formulation inherently lacks temporal directionality. This means the isothermal Jeans model invariably solves for density profiles that best approximate an isothermal sphere configuration at any given instant -- even during core collapse. By mirroring the high-density solution about $t_{\rm merge}$, we essentially reconstruct the temporal sequence of this evolutionary phase while recovering the proper chronological ordering.

In this work, we adopt the ``mirror'' strategy to calculate density profile of SIDM halo in core-collapse state. For a given CDM halo with known $M_{200}$ and $c_{200}$, we first compute the time $t_{\rm merge}\cdot\sigma_m$, in units of ${\rm Gyr\cdot cm^2/g}$, when the high-density and low-density solutions merge. Then, by progressively increasing $t_{\rm age}\cdot\sigma_m$ values in Eq.~\ref{eq:r1} and solving for the density profile of the isothermal core, we determine which solution to adopt: if $t_{\rm age}\cdot\sigma_m < t_{\rm merge}\cdot\sigma_m$, the low-density solution is used; otherwise, the mirrored high-density solution is applied. This process is illustrated in Fig.~\ref{fig:evolution} for a halo with $M_{200}=1\times10^{11}~{\rm M_\odot}$, where the $c_{200}$ is calculated using the mass--concentration relation for CDM from \citet{Dutton2014}. 

In the right panel of Fig.~\ref{fig:evolution}, we present the evolutionary scenario of the velocity dispersion, $\nu_0$, combining the low-density solution and the mirrored high-density solution, omitting the original high-density solution for clarity. After its initial increase, $\nu_0$ remains nearly constant for a long time before $t_{\rm merge}$ but increases steadily afterward. This behavior underscores the rationale for using $t_{\rm merge}$ as the starting point for the core-collapse phase in the isothermal Jeans method.

In Fig.~\ref{fig:profiles1} we compare the density and mass profiles of a core-collapse SIDM halo ($M_{200}=10^{11}~{\rm M_\odot}$) with its CDM counterpart, highlighting differences in matter distribution. The corresponding evolutionary stage is marked by a red star in Fig.~\ref{fig:evolution}. The inner region of the SIDM halo develops a steep and compact core with a radius smaller than 1 kpc, while the density profiles of the SIDM and CDM halos converge at $r_1$. The core-collapse-induced inward mass aggregation results in the SIDM halo exhibiting higher densities than the CDM halo within the 0.03 to 3 kpc range. This steeper inner density profile enables the SIDM halo to accumulate more projected mass up to several kpc, with the increase in enclosed projected mass being most pronounced at 1 kiloparsecs. It also leads to a steeper slope of the surface density around 1 kpc, as illustrated in the middle and right panels of Fig.~\ref{fig:profiles1}.

Readers can refer to Section 2 of Z. Jia et al. (2025, in preparation) for more detailed steps on how to solve $t_{\rm merge}$. Furthermore, we find that the semianalytical model introduced above provides a reliable approximation for the evolution of an SIDM halo. A detailed comparison (S. Li et al., in preparation) with a new high-resolution $N$-body simulation by \citet{Fischer2025} shows that our model accurately reproduces the density profile for over 95\% of the simulation's core-collapse evolution and is sufficiently accurate for the purposes of this study.

\subsection{Computing the lensing properties}
\label{sec:2d_properties}
To compare with observations of J0946+1006, we calculate two lensing properties that, following \citet{Minor2021}, have been identified as well constrained through lens modeling: (1) the average logarithmic slope of the surface density profile within 1 kpc, denoted as $\gamma_{\rm 2D}~(1~{\rm kpc})$, and (2) the projected total mass within 1 kpc, $M_{\rm 2D, tot}~(<1~{\rm kpc})$. 

We adopt these two properties from the best-fit results of ``tNFWmult'' model in \citet{Minor2021}, along with model results from \citet{Despali2024}, as observational benchmarks for comparison with our SIDM model results. To assess whether the modeled SIDM halos produce observational signatures akin to those detected, and to determine the evolutionary stages at which these occur, we computed their complete time evolution and determined the density profiles at specific snapshots along their evolutionary paths, as illustrated by the black dots in Fig.~\ref{fig:evolution}.  

The surface density profile $\Sigma(R)$ is derived from the 3D density profile $\rho(r)$ by integrating along the $z$-axis, with the integration limit $z_{\rm lim}=3 \times r_{200}$:
\begin{equation}
    \Sigma(R)=\int^{z_{\rm lim}}_{-z_{\rm lim}}\rho(\sqrt{R^2+z^2})\mathrm{d}z.
\end{equation}
To measure $\gamma_{\rm 2D}$ at 1~kpc, we fit the surface mass density profile $\Sigma(R)$ with a power-law function $\Sigma(R) = \Sigma_0 R^{\gamma}$ over the radial range 0.75--1.25~kpc, matching the analysis region of \citet{Minor2021}. $M_{\rm 2D}~(< 1~{\rm kpc})$ is then obtained by integrating $\Sigma(R)$ from 0.001 to 1 kpc along $R$.

\begin{figure}
    \centering
	\includegraphics[width=\columnwidth]{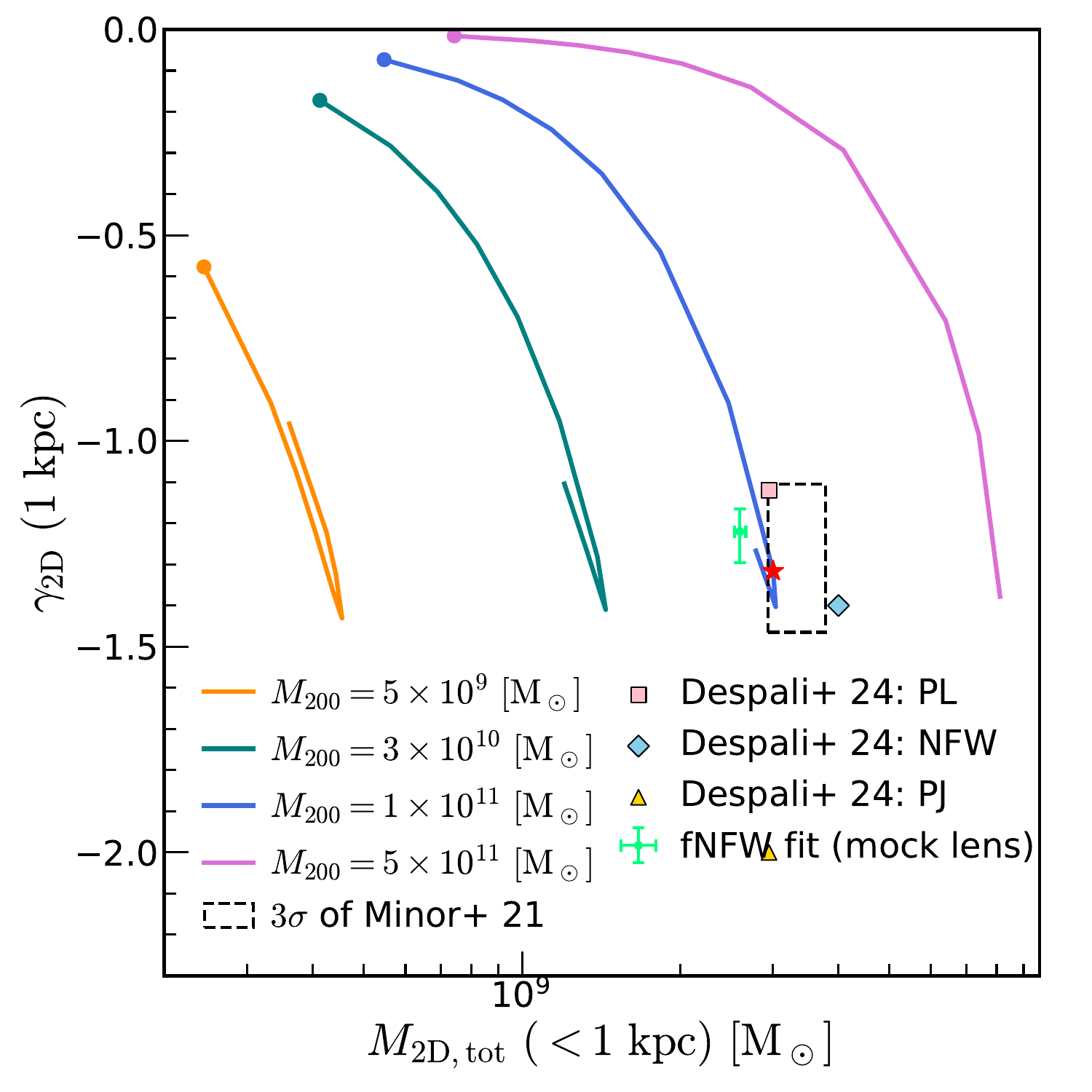}
    \caption{Compare the results of constructed SIDM halos with J0946+1006 observations in the parameter space defined by $\gamma_{\rm 2D}~{\rm (1~kpc)}$ and $M_{\rm 2D, tot}~{\rm (<1~kpc)}$. The evolutionary trajectories of SIDM halos with different masses are represented by color-coded curves, where dots indicate the starting points of post-$t_{\rm merge}$ evolution. The black dashed square shows the 3$\sigma$ measurement uncertainty of \citet{Minor2021}. For the SIDM halo with $M_{200} = 10^{11}~{\rm M_\odot}$, a red star marks the moment when it most closely matches the results from \citet{Minor2021}. Fitting results for different models from \citet{Despali2024} are represented by markers of various shapes. Our tNFW fitting results derived from the mock lensing image are represented by a green point with $1\sigma$ error bars.}
    \label{fig:compare}
\end{figure}

\begin{figure*}
    \centering
	\includegraphics[width=\textwidth]{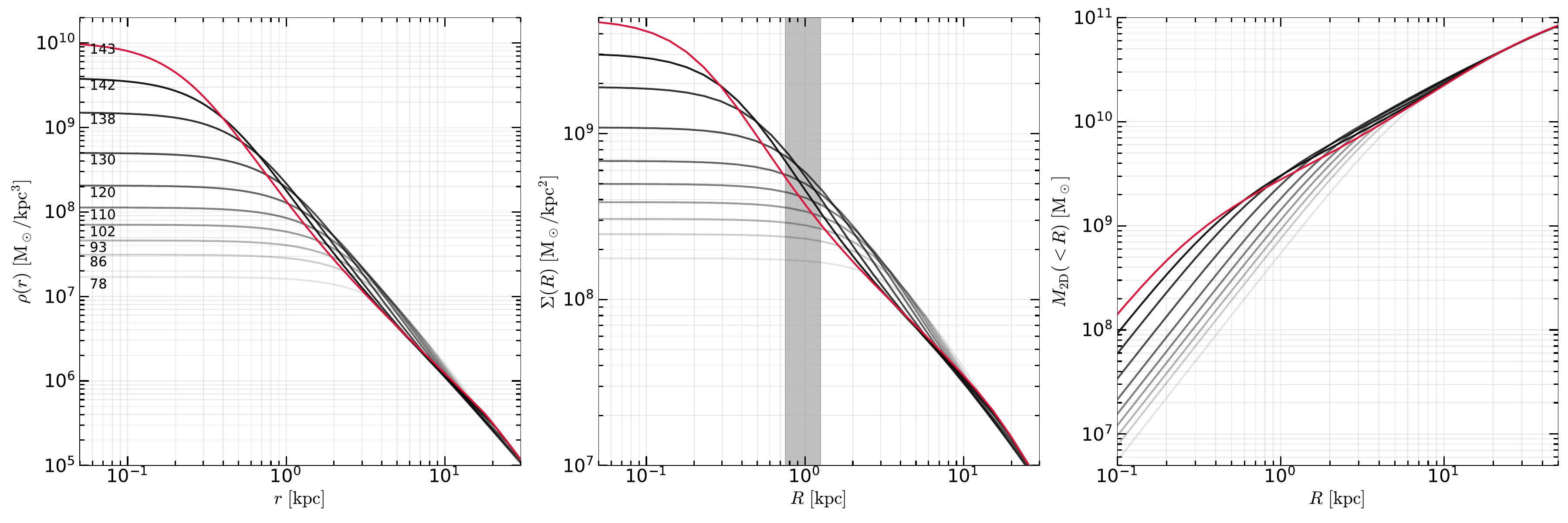}
    \caption{The evolution of the 3D density profiles (left panel), the 2D projected surface density profiles (middle panel), and the 2D enclosed mass profiles (right panel) during the post-$t_{\rm merge}$ stage for an SIDM halo with $M_{200} = 1\times10^{11}~{\rm M_\odot}$. The black curves represent the density profiles prior to the turning point identified in Fig.~\ref{fig:compare}, with line darkness increasing to indicate temporal evolution. The red curve corresponds to the final snapshot after the turning point. In the left panel, the values of $t_{\rm age} \cdot \sigma_m$ (in units of 11.06~${\rm Gyr\cdot cm^2/g}$) are labeled below each line.}
    \label{fig:core-collapse}
\end{figure*}

\section{Results}\label{sec:Results}
\subsection{Comparison with observations}
\label{sec:Results_1}
In Fig.~\ref{fig:compare}, we compare the lensing properties of SIDM halos, evaluated after $t_{\rm merge}$, with those of the observed dark perturber in J0946+1006 \citep{Minor2021, Despali2024}.  We consider four halo masses in ascending order: $5 \times 10^9$, $3 \times 10^{10}$, $1 \times 10^{11}$, and $5 \times 10^{11}~M_\odot$. Their $c_{200}$ values are calculated using the mass--concentration relation from \citet{Dutton2014}.

Our analysis indicates that, across all four selected halo masses, the matter condensation resulting from core collapse causes $\gamma_{\rm 2D}~(1~{\rm kpc})$ to reach approximately -1.4 at its steepest point, which is steeper than the singular isothermal sphere case, -1. This steepening brings the slope close to the value reported by \citet{Minor2021}: $\gamma_{\rm 2D}~(1~{\rm kpc}) = -1.27^{+0.11}_{-0.13}$. However, only the core-collapse halo with a mass of $10^{11}~{\rm M_\odot}$ achieves an enclosed mass comparable to the observed value of $(3.3\pm0.3)\times10^9~{\rm M_\odot}$. Halos with masses significantly larger or smaller than the optimal range accumulate either too much or too little mass in this phase, inconsistent with the observation constraints. 

One can also find that the evolutionary trajectories of the SIDM halos in Fig.~\ref{fig:compare} exhibit a turning behavior, with the degree of turning becoming more pronounced for halos with smaller masses. To examine what happens when an SIDM halo evolves to the end point of the core-collapse phase, Fig.~\ref{fig:core-collapse} shows the evolution of the mass distribution for a halo with $M_{200} = 1\times 10^{11}~{\rm M_\odot}$ during the post-$t_{\rm merge}$ stage. Different lines correspond to different snapshots (marked by black dots in Fig.~\ref{fig:evolution}), with the corresponding $t_{\rm age} \cdot \sigma_m$ values labeled below each line. The red curve represents the end point of its evolutionary trajectory occurring right after the turning point. It can be observed that, after the turning point, the steepest parts of the surface density profile shift to a region smaller than 1 kpc. As a result, the profile slope $\gamma_{\rm 2D}~{\rm (1~kpc)}$ decreases instead.

\begin{figure*}
\centering
    \begin{minipage}[t]{0.4\textwidth}
    \centering
        \includegraphics[width=\textwidth]{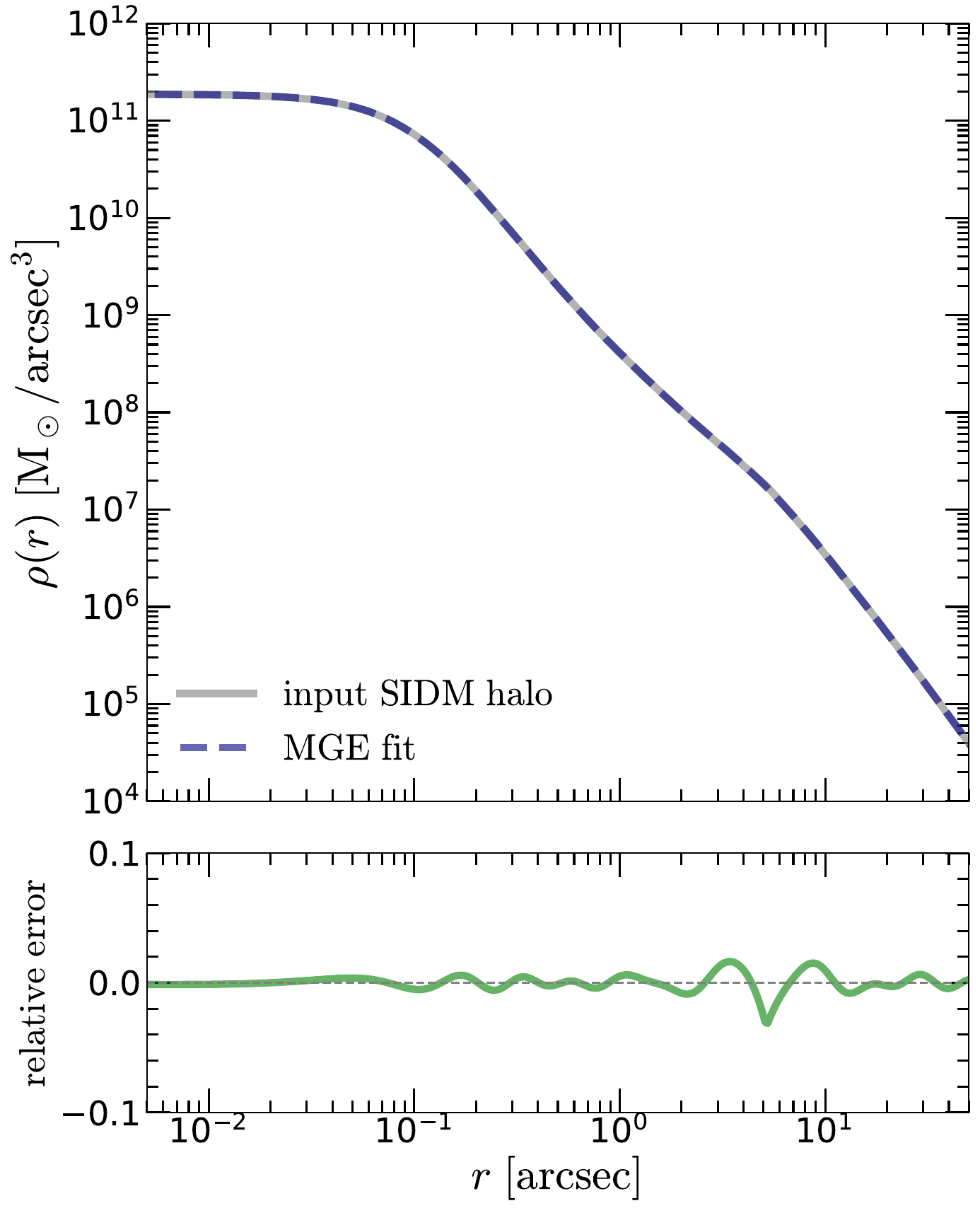}
    \end{minipage}
    \hspace{0.01\textwidth}
    \begin{minipage}[t]{0.54\textwidth}
    \centering
        \includegraphics[width=\textwidth]{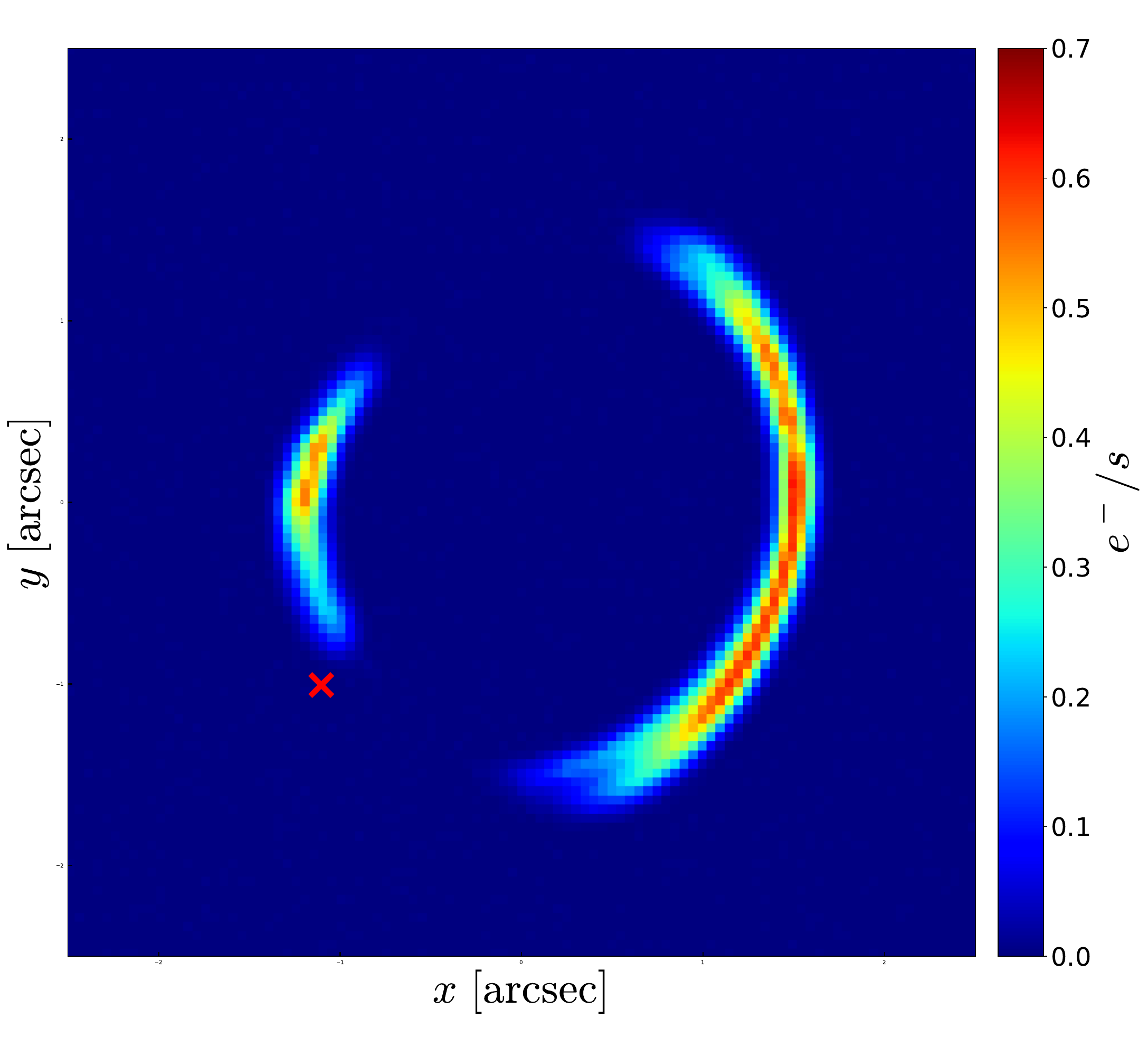}
    \end{minipage}    
    \caption{Left: the MGE fitting result for the density profile of the input SIDM halo. The 3D density profile of the perturber is shown as a gray line, while the MGE fitting result is represented by a navy dashed line. The relative error is illustrated in the bottom subplot. Right: the mock lensing image, with the red cross representing the position of the input SIDM halo.}
    \label{fig:SIDM_MGE_fit}
\end{figure*}

Another important observation is that, for a given halo mass, $M_{\rm 2D}~{\rm (<1~kpc)}$ gradually converges to a finite value during the core-collapse stage, and after $\gamma_{\rm 2D}~{\rm (1~kpc)}$ reaches its turning point, it reverses accordingly. This is because the dark matter self-interaction is only effective at the inner region of the halo, and the redistributable mass during core collapse is constrained by the total mass of the dark halo within $r_1$. Therefore, for any given CDM halo mass, the properties of the density core created by SIDM are not arbitrary; they depend on the initial density profile of the halo at the beginning of its evolution.

In Fig.~\ref{fig:compare}, the 3$\sigma$ uncertainty range reported by \citet{Minor2021} is delineated by a black dashed box. We identify the evolutionary stage of the $M_{200} = 10^{11}~{\rm M_\odot}$ SIDM halo that best matches the observational constraints from \citet{Minor2021}, marking it with a red star in the figure. If we set $t_{\rm age}$ to the cosmic age at the redshift of J0946+1006 ($z$=0.222), we can derive a lower limit for the SIDM cross section, $\sigma_m = 138~{\rm cm^2/g}$, below which no dark matter halo can condense enough to match the observations.

From another perspective, the degree of core collapse in the SIDM halo can be analyzed within the gravothermal fluid framework. In this context, radial heat transport, driven by dark matter self-interactions, can be characterized by the ratio of the scattering mean free path to the halo's dynamical scale height. The time at which the innermost shell of an SIDM halo transitions to the short mean-free-path regime signifies the onset of post-self-similar collapse. At this point, the inner density begins to increase dramatically, making this time a defining collapse timescale, denoted as $t_{\rm c}$. \citet{Essig2019} performed numerical simulations and obtained numerical templates between the core-collapse timescale and other model parameters,
\begin{equation}
    t_{\rm c} = \frac{150}{\beta}\frac{1}{r_s\rho_s\sigma_m}\frac{1}{\sqrt{4\pi G\rho_s}},
\end{equation}
where $\beta=0.75$ is calibrated with $N$-body simulations. Based on this scaling relation, we can transform the product $t_{\rm age}\cdot\sigma_m$ into a dimensionless time variable $\tau=t/t_{\rm c}$. This allows us to analyze the degree of core collapse for the points of interest along the SIDM halo evolution trajectory in Fig.~\ref{fig:compare}. For the SIDM halo with $M_{200} = 10^{11}~{\rm M_\odot}$, we find that our results align most closely with \citet{Minor2021} at $\tau \simeq 1.84$. This indicates that an SIDM halo must evolve to be sufficiently core-collapsed to achieve results consistent with observations.

\subsection{Mock lensing image test}
\label{sec:Results_2}
In addition to comparing the compact, core-collapse SIDM halo with observations, we also explore whether the overconcentrated halos observed in previous strong-lensing studies could be explained by fitting a core-collapse SIDM halo with a CDM halo profile. 

To begin, we simulate a mock lensing image as a mimic of J0946+1006, as shown in Fig.~\ref{fig:SIDM_MGE_fit}. The macro model, which incorporates the mass and light distributions of the main lens and background source, along with their parameters, is based on the fitting results from \citet{He_2024}. The lens mass is modeled as an elliptical power-law (EPL) distribution. As shown in \citet{He_2024}, the lens light is well represented by a multi-Gaussian expansion \citep[MGE;][]{Cappellari_2002}, so we exclude it from our mock image. For the background source, instead of using a parametric model, we adopt a pixelized flux distribution from \citet{He_2024}.

The perturber is modeled as the SIDM halo that best matches the results of \citet[][represented by the red star in Fig.~\ref{fig:compare}]{Minor2021}. Directly calculating the deflection angle of a 3D density profile is challenging. To overcome this issue, we use the MGE model to fit the 3D density profile and then convert the fitting result into a surface density to calculate the perturbation. For a single Gaussian model, the conversion from a 3D profile to a 2D projected profile can be performed analytically with:
\begin{equation}
\Sigma(r;\sigma) = \sqrt{2\pi}\sigma\rho(r;\sigma),
\end{equation}
where $\rho(r;\sigma)\propto\exp[-r^2/2\sigma^2]$ is the 3D Gaussian density profile and $\sigma$ is the standard deviation of the profile. As shown in the right panel of Fig.~\ref{fig:SIDM_MGE_fit}, the MGE model accurately reproduces the density profile of the input SIDM halo, with a relative error of $\lesssim3\%$ within 50 arcsec. The input position of the perturber, marked by a red cross in the right panel, is determined from the power-law fitting results of \citet{Nightingale_24}.

To calibrate the mock image with observational data, we adopt a pixel scale of 0.05$\arcsec$ and convolve it with a Gaussian PSF ($\sigma$ = 0.05\arcsec) to closely match the image quality of SLACS lensing systems. The sky background is set to $0.1~{\rm e^-/(pixel\cdot s)}$, consistent with an HST F814W observation of the extragalactic sky. We adjusted the exposure time to achieve a signal-to-noise ratio of approximately 100 for the brightest pixel in the lensing arc, which is comparable to the observations of SLACS lenses.

Next, we fit the simulated mock lensing image. The lens mass is modeled using an EPL profile plus external shear. Following \citet{Nightingale_24}, we model the source light using a Voronoi mesh with natural neighbor interpolation \citep{Sibson1981}. We smooth the source reconstruction by applying adaptive regularization based on the source fluxes at each point. The regularization formalism is the same as \citet{Suyu2006} and the likelihood function could be found in Eq.~17 of \citet{Nightingale_24}. The positions of the Voronoi mesh centers on the image plane are adjusted to follow the morphology of the lensing arc, with more source pixels reconstructing the brighter regions. A detailed description of the adjustment scheme can be found in \citet{Wang_2025}. 

For the subhalo, we adopt a tNFW profile,
\begin{equation}
    \rho(r)=\frac{\rho_s}{(r/r_s)(1+r/r_s)^2} \frac{r_t^2}{r_t^2+r^2}, 
\end{equation}
where $\rho_s$ and $r_s$ are the characteristic density and scale radius of the NFW profile (Eq.~\ref{NFW}), and $r_t$ is the truncation radius. We fit the mock lensing image using the automated Source, Lens, and Mass (SLaM) strong lens modeling pipeline, which is implemented in the open-source software \texttt{PyAutolens} \citep{slam_Cao_2022, slam_Amy22, slam_he, He_2024, Nightingale_24}. The model takes as input three parameters: the halo mass ($M_{200}$), concentration ($c_{200}$), and truncation radius ($r_t$), while $\rho_s$ and $r_s$ are subsequently derived from $M_{200}$ and $c_{200}$. We use the nested samplers \texttt{Dynesty} \citep{dynesty} and \texttt{Nautilus} \citep{nautilus} to obtain the posteriors. 

\begin{figure}
	\includegraphics[width=\columnwidth]{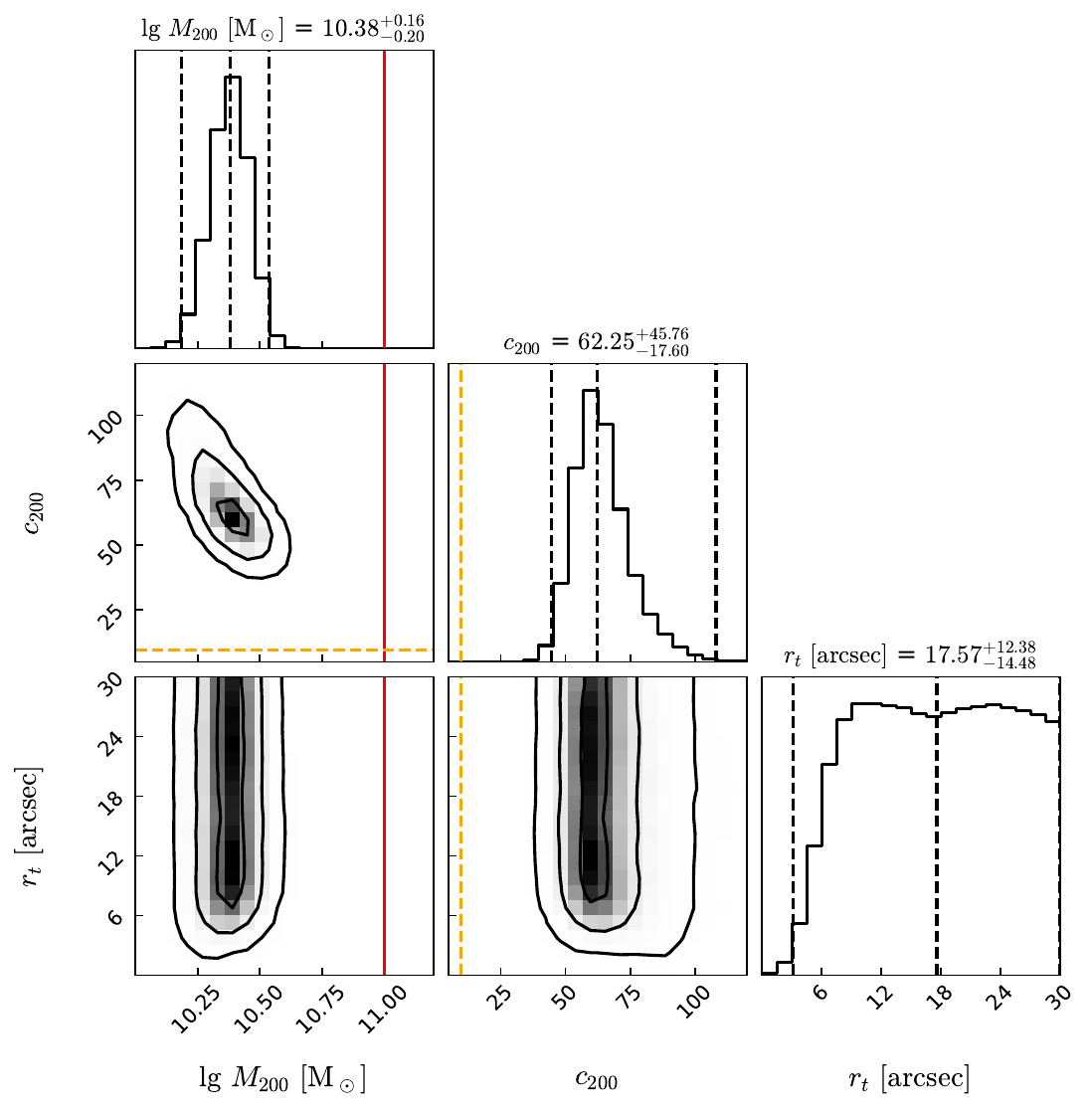}
    \caption{The posterior distributions of $M_{200}$, $c_{200}$ and $r_t$ from the tNFW model fitting. In the 1D posteriors, the black dashed lines indicate the medians and the $3\sigma$ boundaries. In the 2D posterior, the black contours represent the $1\sigma$, $2\sigma$, and $3\sigma$ regions. The red line marks the $M_{200}$ value of the input SIDM halo used to generate the mock image, while the orange dashed line corresponds to its $c_{200}$ value.}
    \label{fig:triangle}
\end{figure}

\begin{figure*}
	\includegraphics[width=\textwidth]{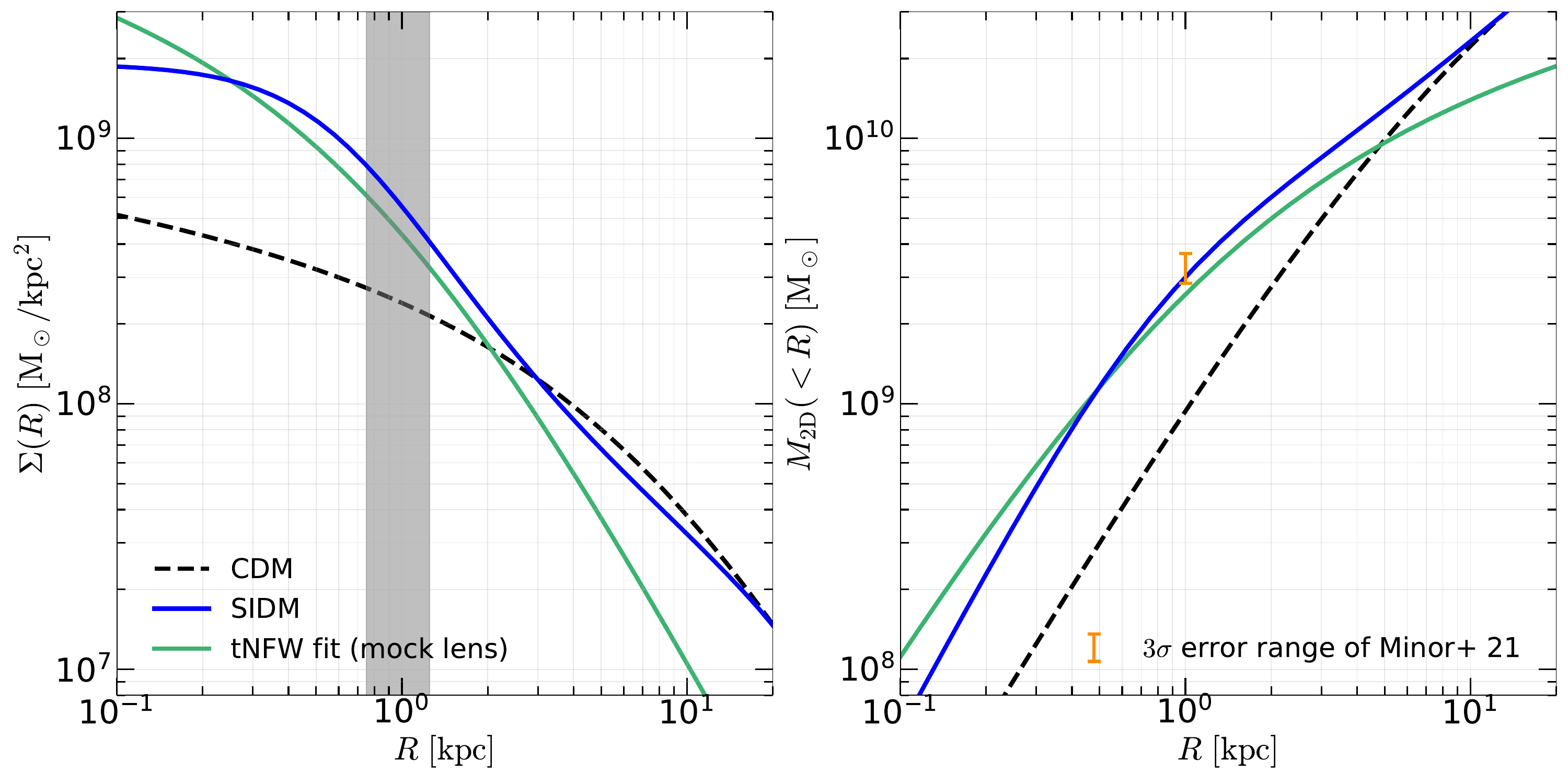}
    \caption{Same format as Fig.~\ref{fig:profiles1}, but here we include the best-fit tNFW profile from our mock lensing image, with $M_{200}=2.40\times10^{10}~{\rm M_\odot}$ and $c_{200}=62.25$. Left: 2D projected surface density profiles, with the gray band indicating the radius range used to calculate the log slope, $\gamma_{\rm 2D}$. Right: 2D enclosed mass profiles. The error bar indicates the $3\sigma$ measurement uncertainty of $M_{\rm 2D, tot}~{\rm (<1~kpc)}$ from \citet{Minor2021}.}
    \label{fig:profiles2}
\end{figure*}

During the fitting process, the macro model is initially fitted without the perturber. We first use a parametric lens and source model to obtain an initial parameter estimate, then refine the fit with a pixelized source model, sampling parameters around the previous best result. This iterative approach ensures the global best solution. A detailed overview is provided in Appendix~\ref{sec:appendix_macro_model_fitting}.

After completing the macro model fitting, we introduce a tNFW perturber and refit the image. In this step, the priors for the main lens model parameters are set as Gaussian distributions centered on the best-fit values from the macro model. The pixelized source model parameters are fixed, except for the fluxes, which are solved linearly.

In Fig.~\ref{fig:triangle}, we show the joint posterior distribution of $M_{200}$, $c_{200}$ and $r_t$ derived from the tNFW model fitting. The best-fit tNFW halo has a halo mass systematically lower than that of the input SIDM halo. Supposing that the \(1\sigma\) scatter of concentration is 0.15 dex \citep{Wang2020}, the inferred concentration is extremely high, exceeding the CDM mass--concentration relation by \(5.43\sigma\). This outcome is consistent with previous analyses of J0946+1006 \citep{Minor2021, Ballard_24, Enzi2024, Despali2024}. Our results indicate that a model misspecification--incorrectly modeling a core-collapse SIDM subhalo as a tNFW halo--leads to an overconcentrated solution. 

Note that, since we do not know whether or not the dark perturber has undergone tidal stripping, we create the mock image using an untruncated SIDM halo as the input perturber. In practice, truncation has negligible impact on both the subhalo's lensing effect (which primarily originates from the core region) and its inferred $M_{200}$ (which is essentially an extrapolation of the inner profile). As shown in Fig.~\ref{fig:triangle}, the observational data places no constraints on the tidal radius. Further discussion of the effects of tidal stripping can be found in Section~\ref{sec:Discussion3}.

For each sampled tNFW profile, we calculate its $\gamma_{\rm 2D}~{\rm (1~kpc)}$ and $M_{\rm 2D}~{\rm (<1~kpc)}$. We then compute the median and $1\sigma$ scatter of these 2D parameters and plot them in Fig.~\ref{fig:compare}. One can find that the best-fit tNFW profile falls slightly outside the \(3\sigma\) region around the observed values. Fig.~\ref{fig:profiles2} further shows that, compared to the input SIDM halo, the inferred tNFW profile has a slightly flatter \(\gamma_{\rm 2D}\) around 1 kpc and a lower \(M_{\rm 2D}\) within 1 kpc.  Although some differences exist, the tNFW profile is still able to capture the lensing effects produced by a core-collapse halo. However, the fitting process tends to yield a lower halo mass than the true value while significantly overestimating the concentration parameter.

\section{Discussion}\label{sec:Discussion}
\subsection{Galaxies in the halo}
\label{sec:Discussion1}

\begin{figure}
\centering
	\includegraphics[width=\columnwidth]{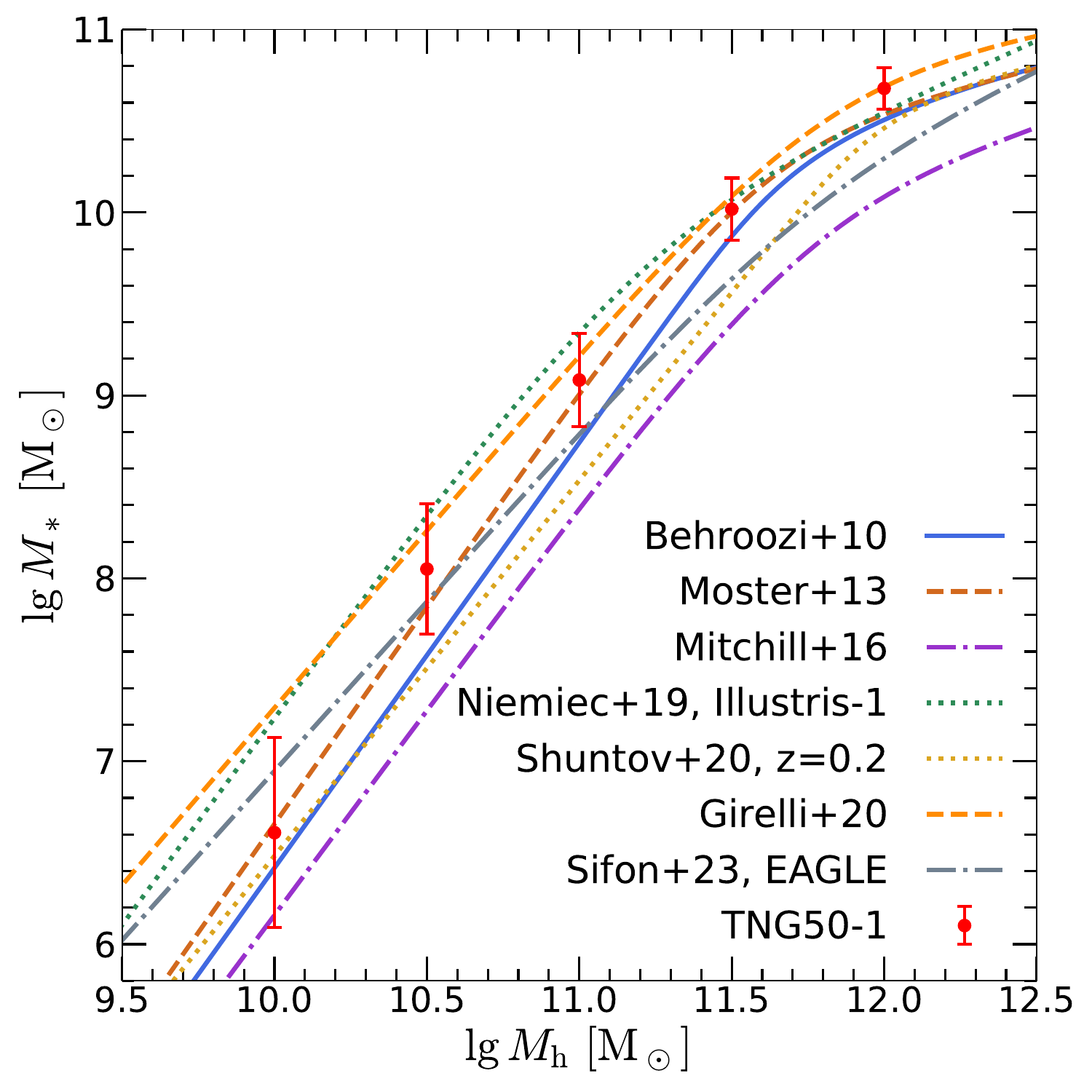}
    \caption{Median SHMRs from different studies. Colored curves show best-fit relations from \citet{Behroozi2010}, \citet{Moster2012}, \citet{Mitchell2015}, \citet{smhr_Niemiec_2019}, \citet{Girelli2020}, \citet{Shuntov_2022}, and \citet{Sifon2024}, while the TNG-50 simulated result is marked by red dots with $1\sigma$ error bars.}
    \label{fig:shmr}
\end{figure}

In \citet{Despali2024}, it is argued that if a dark matter halo with a mass greater than $10^{10}~{\rm M_\odot}$ exists in J0946+1006, it should already have been detectable in current observations. Although constraining the stellar-to-halo mass relation (SHMR) for SIDM would require full hydrodynamic cosmological simulations, we adopt the reasonable assumption that SIDM follows an SHMR similar to CDM. This allows us to estimate the stellar mass of J0946+1006 if its lensing signal originates from an SIDM halo.

In Fig.~\ref{fig:shmr}, we present the SHMR for field halos from various observational and theoretical studies \citep{Behroozi2010, Moster2012, Mitchell2015, smhr_Niemiec_2019, Girelli2020, Shuntov_2022, Sifon2024}, a choice motivated by our modeling of the perturber as a complete NFW halo. For the TNG-50 simulation \citep{TNG50-1, TNG50-2}, we select halos with masses of $10^{10\pm0.1}$, $10^{10.5\pm0.1}$, $10^{11\pm0.1}$, $10^{11.5\pm0.1}$, and $10^{12\pm0.1}~{\rm M_\odot}$ and calculate the stellar mass distributions of their central galaxies. While the SHMR estimates differ by up to an order of magnitude across different works, they consistently indicate that a dark matter halo with a mass of $10^{11}~{\rm M_\odot}$ is expected to host a galaxy with a stellar mass in the range of $10^8$–$10^9~{\rm M_\odot}$. Analysis by \citet{Despali2024} based on the Illustris-TNG simulations further supports this result, showing that the mass of the satellite galaxy hosted by the subhalo in J0946+1006 should -- given the mass of the subhalo -- be larger than $10^{8.5}~{\rm M_\odot}$.

Thus, a massive SIDM halo is expected to host a luminous galaxy, creating a potential tension with the nondetection of a stellar counterpart for the perturber. The initial study by \citet{Vegetti_2010} established a stringent \(3\sigma\) upper limit of $5\times10^6~{\rm L_\odot}$. This limit was derived indirectly by analyzing the residual image -- after subtracting a best-fit lens model that assumed a dark subhalo -- and measuring the statistical noise within a defined aperture at the perturber's location. A galaxy with a stellar mass of $10^8~{\rm M_\odot}$ remaining below this limit would imply an extraordinarily high mass-to-light ratio of \(\sim 20\), which is challenging under normal galaxy formation scenarios.

However, the observational constraints on this counterpart are currently uncertain. Recognizing that the residual-based method risks underestimating the brightness, \citet{Minor2021} revisited the constraint with a direct modeling approach. They incorporated a Gaussian light profile for the subhalo into their lens model and fitted its luminosity and size simultaneously with all other parameters. This yielded a more conservative upper bound of $1.2\times10^8~{\rm L_\odot}$, which the authors cautiously note may be an overestimation due to potential contamination from unsubtracted foreground light.

Moreover, a recent reanalysis of this strong-lensing system by \citet{He2025} using \texttt{PyAutoLens} showed that the observed perturbation is better explained by a luminous dwarf satellite with a luminosity of $2.5^{+0.7}_{-0.9}\times10^8\ {\rm L_\odot}$ and an NFW halo whose properties are consistent with CDM predictions. However, this statistical preference should be interpreted carefully, as the current observational data are insufficient to unambiguously distinguish this scenario from that of a very compact, purely dark halo.

The significant discrepancy among these observational analyses leaves the perturber's true luminosity as a pivotal and unresolved issue, thereby preventing a conclusive test of the core-collapse SIDM halo scenario. Ultimately, its viability hinges on future observations sensitive enough to firmly establish the presence or absence of a faint host galaxy.

\subsection{Effect of mass--concentration relation}
\label{sec:Discussion2}
While our fiducial SIDM halos adopt median concentrations based on the \(c_{200}(M_{200}, z)\) relation from \citet{Dutton2014}, it is necessary to consider scenarios where J0946+1006 significantly exceeds this median value. This is because, if a halo exhibits higher concentration before core collapse, its core region would accumulate more mass, subsequently leading to a core-collapse central region with greater mass. As a result, lower-mass halos with higher concentrations could produce observational effects similar to those of our fiducial halo. Note that this phenomenon has already been discussed in the background of CDM perturbers \citep{Amorisco2022}. In \citet{Nadler2023}'s $N$-body simulation, J0946+1006-like SIDM subhalos also showed enhanced concentrations compared to the field population.

\begin{figure}
\centering
	\includegraphics[width=\columnwidth]{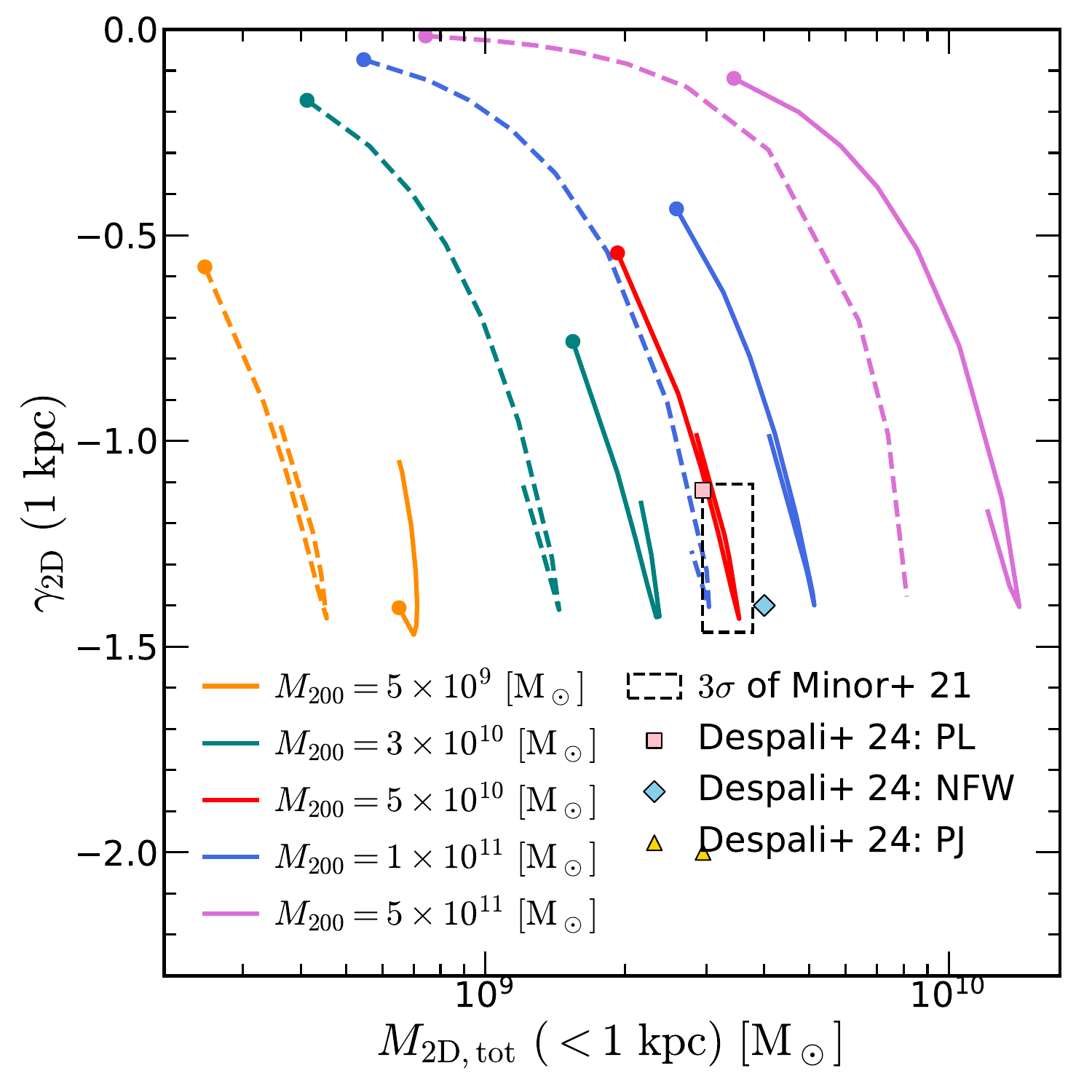}
    \caption{Same format as Fig.~\ref{fig:compare}: dashed curves show the original evolutionary tracks, while solid curves display the evolution of an SIDM halo with precollapse $c_{200}$ higher than the median by $3\sigma$ ($=0.45$ dex).}
    \label{fig:compare_high_c}
\end{figure}

Assuming a \(1\sigma\) scatter in concentration of 0.15 dex at fixed halo mass and redshift, we generate SIDM halos with \(M_{200}\) values spanning the range shown in Fig.~\ref{fig:compare}. For each halo, we set the concentration \(c_{200}\) to be \(3\sigma\) above the median value given by the concentration–mass relation (i.e. 0.45 dex higher). Compared to halos with median \(c_{200}\), we find that a \(3\sigma\) concentration enhancement shortens the SIDM evolution timescale \(t_{\rm merge}\) by a factor of approximately 18 and increases the central density \(\rho_0\) by an order of magnitude.


\begin{figure*}
\centering
	\includegraphics[width=\textwidth]{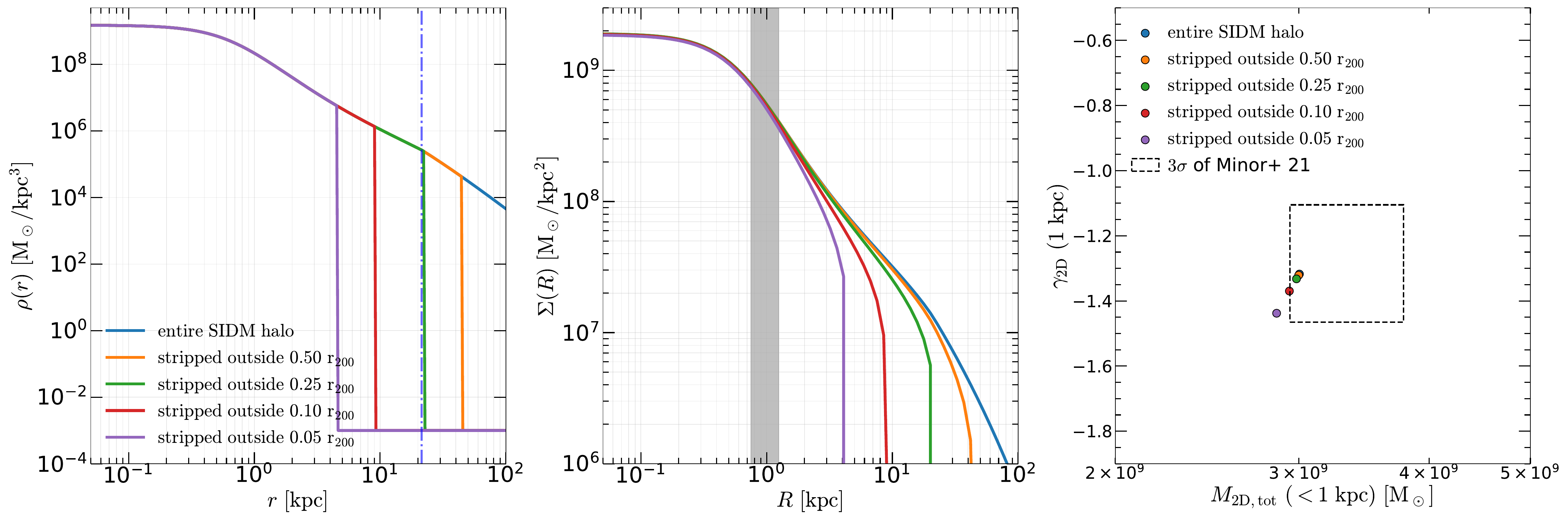}
    \caption{Left and middle: Same format as Fig.~\ref{fig:profiles1}, but showing the core-collapse SIDM halo with color-coded tidal stripping degrees. Right: same format as Fig.~\ref{fig:compare} but comparing the observationally matched SIDM halo across different tidal stripping degrees.}
    \label{fig:stripped_profile}
\end{figure*}
The 2D lensing parameters of a high-concentration halo that is undergoing core collapse are compared to those of a fiducial halo in Fig.~\ref{fig:compare_high_c}. Observed data are well explained by a subhalo of lower mass ($M_{200}=5\times10^{10}~{\rm M_\odot}$),  but high concentration ($c_{200}$ above the median by $3\sigma$). However, according to the SHMR in Fig.~\ref{fig:shmr}, even this lower-mass halo should host a galaxy with mass $5\times10^{7}~{\rm M_\odot}$, and whether this is consistent with observations still remains an unresolved tension due to the current uncertainty in the subhalo's luminosity.

\subsection{Tidal truncation of the dark perturber}
\label{sec:Discussion3}
So far, we assume the dark halo follows an NFW profile to facilitate comparison with theoretical predictions of the SHMR and to estimate the stellar mass of the galaxy it should host. However, if the perturber resides within the host dark matter halo of the lens galaxy, it would inevitably experience tidal stripping. Given that the dark halo required to explain the observational data is highly compact, we estimate that tidal stripping would predominantly remove the halo’s outer regions while leaving its inner profile largely intact.

In left and middle panels of Fig.~\ref{fig:stripped_profile}, we remove mass outside different radial ranges of the SIDM halo that best matches the result of \citet[][represented by the red star in Fig.~\ref{fig:compare}]{Minor2021}. We then compute the 2D projected density profiles and 2D enclosed mass profiles and compare the lensing properties, as shown in the right panel, with those of the entire SIDM halo. Our theoretical calculations reveal that for a compact core-collapse SIDM halo, even if more than 95\% of the total mass is stripped (corresponding to the mass outside 0.05 $r_{200}$ being removed), $M_{\rm 2D}~{\rm (<1~kpc)}$ would decrease by merely 4.9\%, and $\gamma_{\rm 2D}~{\rm (1~kpc)}$ would increase by only 9.1\%. This indicates that the core properties that produce strong-lensing effects remain robust against tidal stripping. 

Note that in the above analysis, we assume the subhalo has already entered the core-collapse phase before falling into the main halo. In reality, if core collapse occurs after the subhalo falls into the main halo, the environment within the main halo may accelerate the core-collapse process. A detailed analysis of how this process impacts core collapse would require considering heat conduction, which cannot be handled using the isothermal method. However, in principle, tidal stripping will not affect the mass in the core-collapse region of the halo \citep{Nishikawa2020}; therefore, we do not expect that this will impact the main conclusion of this study. \citet{Nadler2023} also found in their $N$-body simulations that core-collapse halos are more resilient to tidal disruption by their host halos, further supporting our assumption.

\section{Conclusions}
\label{sec:Conclusions}
In this work, we conduct a detailed investigation into whether the SIDM scenario can explain the very compact dark perturber previously identified in SDSS J0946+1006. We find that while a core-collapse SIDM halo is a viable candidate for the observed perturber, it imposes strict mass requirements. Specifically, only an SIDM halo with \(M_{200} \approx 1 \times 10^{11} \, {\rm M_\odot}\) aligns well with the observational constraints for the perturber in SDSS J0946+1006. Lower-mass halos, even if they undergo core collapse, fail to accumulate sufficient mass within the central 1 kpc region to match the observed properties.  

By setting the formation time of a \(10^{11} \, {\rm M_\odot}\) halo equal to the cosmic time, we derive a lower bound on the SIDM cross section of \(\sigma_m = 138 \, {\rm cm^2/g}\). This conclusion is largely unaffected by tidal stripping effects on the halo.  

Furthermore, we investigate whether fitting an SIDM halo with an NFW-based model could lead to misinterpretations of its mass distribution. We generate a mock lensing image of SDSS J0946+1006 using the above SIDM profile for the perturber and fit it with a tNFW profile. The resulting best-fit \(M_{200}\) is 3 times lower than the true SIDM halo mass, while the inferred concentration, \(c_{200} = 62.25\), is significantly overestimated, deviating by \(>5\sigma\) from the CDM mass--concentration relation. This outcome is broadly consistent with the findings of \citet{Minor2021}.  

However, although a core-collapse SIDM halo can explain the observed perturbation, the absence of a visible galaxy within such a massive halo requires careful consideration of its expected luminosity. For a halo of $1\times10^{11}~{\rm M_\odot}$, CDM's standard SHMRs predict a central galaxy with $M_*\sim10^8-10^9~{\rm M_\odot}$. Even with a $3\sigma$ increase in $c_{200}$ lowering the required SIDM halo mass to $5\times10^{10}~{\rm M_\odot}$, a host galaxy with a stellar mass $>5\times10^7~{\rm M_\odot}$ is still expected. This creates a potential tension with the nondetection of a luminous counterpart, an issue that cannot be fully resolved without a definitive measurement of the perturber's true luminosity.

Future high-sensitivity follow-up observations with the James Webb Space Telescope (JWST) are required to place more precise constraints on the perturber's mass distribution and definitively determine whether it is associated with a faint host galaxy. Additionally, upcoming space-based surveys, such as those conducted by Euclid, the China Space Station Telescope, and the Roman Space Telescope, will reveal whether such compact dark perturbers are common. This combined approach of detailed characterization and statistical census will be crucial for reshaping our understanding of dark matter.

\begin{acknowledgments}
We thank Daneng Yang, Hai-Bo Yu, Moritz Fischer, and Simon White for helpful discussions. S.L., R.L., K.W., X.C., and X.M. acknowledge the support by National Key R\&D Program of China No.~2022YFF0503403, the support of the National Nature Science Foundation of China (No.~11988101), the support from the Ministry of Science and Technology of China (No.~2020SKA0110100),  the science research grants from the China Manned Space Project, CAS Project for Young Scientists in Basic Research (No.~YSBR-062), and the support from K. C. Wong Education Foundation. Z.J. and F.J. acknowledge the support of the National Natural Science Foundation of China (No.~12473007) and the Beijing Natural Science Foundation (QY23018). 
\end{acknowledgments}

\appendix
\section{Solving the Poisson equation}
\label{sec:appendix_poisson_equation}
Not only is the influence of baryonic matter reflected in the change of the SIDM halo's density distribution through solving the Poisson equation (Eq.~\ref{eq:PoissonEquation}), but also the gradual growth of the baryonic potential causes adiabatic contraction in the dark halo, leading to a more centrally concentrated distribution of dark matter \citep[e.g.,][]{Gnedin2004}. One advantage of the \textit{outside-in} approach is its flexibility in incorporating baryonic effects on the dark halo's adiabatic contraction. In the code of \citet{Jiang2023}, this is achieved by first contracting the CDM halo using the baryonic potential \citep[based on the model of][]{Gnedin2004} to compute $r_1$ and then stitching the contracted halo with the inner isothermal core.

To facilitate solving the Poisson equation (Eq.~\ref{eq:PoissonEquation}) and calculating the adiabatic contraction effect of baryons on the halo, a common approach is to model the baryon density profile using a Hernquist profile \citep{Hernquist},
\begin{equation}
    \rho_{\rm b}(r) = \frac{\rho_{\rm b0}}{\frac{r}{r_0}\left(1+\frac{r}{r_0}\right)^3},
	\label{eq:Hernquist}
\end{equation}
where $\rho_{\rm b0}$ is determined by the baryon mass $\rho_{\rm b0} = \frac{M_{\rm b}}{2\pi r_0^3}$ and $r_0$ is the scale radius. After substituting the expressions for $\rho_{\rm b}$ and $\rho_{\rm iso}$, Eq.~\ref{eq:PoissonEquation} becomes
\begin{equation}
    \frac{1}{r^2} \frac{\mathrm{d}}{\mathrm{d}r} \left( r^2 \frac{\mathrm{d}\Phi}{\mathrm{d}r} \right) = 4\pi G \rho_{\rm iso0} \exp\left(-\frac{\Delta\Phi}{\nu_0^2}\right) + \frac{4\pi G \rho_{\rm b0}}{\frac{r}{r_0}\left(1+\frac{r}{r_0}\right)^3}.
\end{equation}
For simplicity, we perform the first variable replacement, $x = r/r_0$ and $\frac{\mathrm{d}}{\mathrm{d}r}=\frac{1}{r_0}\frac{\mathrm{d}}{\mathrm{d}x}$,
\begin{equation}
    \frac{1}{x^2} \frac{\mathrm{d}}{\mathrm{d}x} \left( x^2 \frac{\mathrm{d}\Phi}{\mathrm{d}x} \right) = 4\pi Gr_0^2 \rho_{\rm iso0} \exp\left(-\frac{\Delta\Phi}{\nu_0^2}\right) + \frac{4\pi Gr_0^2\rho_{\rm b0}}{x\left(1+x\right)^3}.
\end{equation}
Furthermore, we perform the second variable replacement, $y = \frac{x}{1+x}$ and $\frac{\mathrm{d}}{\mathrm{d}x}=(1-y)^2\frac{\mathrm{d}}{\mathrm{d}y}$,
\begin{equation}
    \frac{1}{y^2} \frac{\mathrm{d}}{\mathrm{d}y} \left( y^2 \frac{\mathrm{d}\Phi}{\mathrm{d}y} \right) = \frac{4\pi Gr_0^2\rho_{\rm iso0}}{(1-y)^4} \exp\left(-\frac{\Delta\Phi}{\nu_0^2}\right) + \frac{4\pi Gr_0^2\rho_{\rm b0}}{y}.
\end{equation}
Finally, by setting $h(y)=-\frac{\Phi(r)}{\nu_0^2}$ and defining $a\equiv4\pi Gr_0^2\rho_{\rm iso0}/\nu_0^2$ and $b\equiv4\pi Gr_0^2\rho_{\rm b0}/\nu_0^2$, the second-order differential equation can be rewritten in a dimensionless form as
\begin{equation}
    \frac{\mathrm{d}^2h}{\mathrm{d}y^2}+\frac{2}{y}\frac{\mathrm{d}h}{\mathrm{d}y}+\frac{ae^h}{(1-y)^4}+\frac{b}{y}=0.
\end{equation}
With the initial conditions $\Phi(0) = 0$ and $\Phi^\prime(0) = GM_{\rm b}/r_0^2$ \citep{Jiang2023}, this equation can be easily solved numerically.

\section{Macro Model Fitting}
\label{sec:appendix_macro_model_fitting}
We utilize the SLaM pipeline to fit the lensing image with the macro model, which ignores the contribution from potential mass perturbations. An overview of the pipeline structure is as follows.

(a) \textit{Source parametric pipeline}. A parametric lens mass and source light model are used to model the lensing image, providing a rapid initial estimation of the lens's mass. Specifically, the lens mass distribution is described with a singular isothermal ellipsoid plus shear model, while the source light is modeled using multiple Gaussian expansion models that include a set of 20 Gaussians sharing a common center and ellipticities.

(b) \textit{Source pixelized pipeline}. Pixelized source models are employed to capture the complex morphology of the background source and refine the lens mass model provided by the \textit{source parametric pipeline}. Specifically, this pipeline consists of two stages. In the first stage, the lens mass estimation from the \textit{source parametric pipeline} is inherited as the informative Gaussian prior, in combination with a source pixelization scheme that adapts to the lensing magnification and a constant regularization scheme, to refine the lens mass model and provide an initial estimation of the source's (potentially irregular) brightness distribution. In the second stage, the lens mass model is fixed to the estimated values given by the first stage, and the pixelized source model is updated to a more advanced one, where the pixelization and regularization adapt to the source brightness distribution.

(c) \textit{Mass pipeline}. A more general EPL plus shear mass model is used to improve the lens mass estimation given by the \textit{source pixelized pipeline}. The prior settings for the EPL and shear models are inherited from the best lens mass estimation from the \textit{source pixelized pipeline}, except for the EPL's slope, which is assumed to be uniformly distributed between 1.0 and 3.0. For the source model, the pixelized source model from stage 2 of the \textit{source pixelized pipeline} is used, with all the hyperparameter values fixed.

\bibliography{reference}{}
\bibliographystyle{aasjournal}



\end{document}